# CNN-based InSAR Coherence Classification

Subhayan Mukherjee[1], Aaron Zimmer[2], Xinyao Sun[1], Parwant Ghuman[2], Irene Cheng[1]
[1]Department of Computing Science, University of Alberta, Edmonton, Canada
[2]3v Geomatics, Vancouver, Canada

{mukherje, xinyao1, locheng}@ualberta.ca         {azimmer, pghuman}@3vgeomatics.com

*Abstract*—**Interferometric Synthetic Aperture Radar (InSAR) imagery based on microwaves reflected off ground targets is becoming increasingly important in remote sensing for ground movement estimation. However, the reflections are contaminated by noise, which distorts the signal's wrapped phase. Demarcation of image regions based on degree of contamination ("coherence") is an important component of the InSAR processing pipeline. We introduce Convolutional Neural Networks (CNNs) to this problem domain and show their effectiveness in improving coherence-based demarcation and reducing misclassifications in completely incoherent regions through intelligent preprocessing of training data. Quantitative and qualitative comparisons prove superiority of proposed method over three established methods.**

## I. Introduction

In the past few decades, there has been an increasing use of remote sensing using activate microwave in general, and Synthetic Aperture Radar Interferometry (InSAR) in particular. The phase component of InSAR signal encodes the distance between satellite and ground target, and phase unwrapping can help create highly accurate Digital Elevation Maps (DEMs). Unfortunately, the phase data often suffers contamination due to noise arising from numerous sources, e.g. atmospheric factors. Hence, denoising of the phase prior to unwrapping is essential to optimize the InSAR processing pipeline as a whole.

Ground images acquired using InSAR are "non-stationary" due to changes in topography and displacement of ground along the satellite's line of sight. The boxcar filtering approach, which is still widely used today, involves computing moving average using a rectangular window. But the non-stationary nature of InSAR signal adversely affects the performance of sample average methods like boxcar [1]. Also, the strong smoothing effect of boxcar filtering renders spatial resolution loss, in addition to noticeable phase and coherence estimation errors near signal discontinuities. Consequently, various filtering methods have tried to address this problem of estimating non-stationary InSAR phase. These methods are broadly classified as spatial methods, e.g., Lee [2], and frequency-based methods, e.g., Goldstein [3]. Both of these filters, as well as [4] are adaptive to local fringe direction. The original and modified Goldstein filter of Baran et al. [5] preserves the signal in high coherence (low noise) areas, which makes them locally noise-adaptive as well. This emphasizes the importance of accurate classification of image regions based on coherence. Lee filter's enhancements [6-9] improve adaptation to the fringe structure in the local signal neighborhood, whereas modifications to the Goldstein and Baran filters avoid under-filtering the incoherent regions via improved coherence estimation [10, 11]. Other approaches include wavelet domain methods [12] including un-decimated wavelet transform [13] and wavelet packets [14], local modeling based on polynomial approximation [15], sparse coding [16], Markov Random Field (MRF) based methods [17, 18] (though prior distribution modeling required in MRF is difficult) and non-local filtering methods [19-21].

While Neural Network based SAR images despeckling [22-25] and improvement of geo-localization accuracy for optical satellite images [26] have been explored, Convolutional Neural Network (CNN) based learning approaches to InSAR images' coherence classification have not been investigated. This paper proposes coherence classification using a CNN. We show how intelligent MRF-based preprocessing of raw coherence can train the coherence classifier CNN to improve demarcation of coherent and incoherent regions in the input noisy interferogram, and also reduce misclassifications in completely incoherent regions, which is ubiquitous in methods like boxcar.

Our work is distinct from [17, 18] in that we use MRF for thresholding the raw coherence while preparing training labels. In contrast, the authors in [17, 18] use MRF for denoising. Our work is also different from [27-32], as the type of data we use is InSAR, not (Pol)SAR, and our objective is not classifying ground targets, but separating coherent and incoherent regions.

## II. Proposed Method

Fig. 1 shows our coherence classification CNN. We do not need to constrain the network's training image size or that for running inference using the network after training. We can use patches or whole image as input. The CNN has two input channels, representing real and imaginary components of the complex interferogram image. Thus, information from both channels is used simultaneously. First, they are considered separately to saturate outlier amplitudes (some input interferogram pixels might have extremely high amplitudes, which can degrade the training / inference). Let interferogram pixels be $Z = [z_1, z_2, ... z_N]$ having amplitudes $A = [a_1, a_2, ... a_N]$. Amplitude of each pixel can be thresholded as $A' = \text{saturate}(A, M)$, where $M$ is the mask representing outlier amplitudes. The outliers are computed following the method [33]. We saturate and normalize real and imaginary channel values to lie between -1 and +1. We then add 1 to each channel in order to use the Rectified Linear Unit (ReLU) activation for introducing nonlinearity in the CNN to learn complex features.

In Fig.1, each box represents a CNN layer. The integer at the top (1, 16) represents the number of output feature maps, whereas filter dimension is mentioned at the bottom (3×3). Each 2D convolutional layer learns a number of filters. We use separable convolutions to reduce the number of weights to be trained for faster convergence. The output of the last convolutional layer is a single channel, as we need to classify each pixel based on only its coherence's amplitude. The activation for final convolution layer is sigmoid instead of ReLU because, pixel coherence is a value between 0 and 1 (sigmoid output). Binary Cross-entropy between the output channel and the training target is reduced using the popular Adam optimizer to train the network, by updating its filter weights and biases, using gradient backpropagation. The Xavier method [34] is used to initialize the weights of the CNN before training starts. We similarly build another CNN to filter the noisy interferograms. It has an autoencoder structure of 3×3 filters [16-8-Maxpool-8-Upsample-16-2] with ReLU activation even on the last layer. Maxpooling layers subsample their input feature maps by a factor of 3; Upsampling layers bring them back to their original size. The last convolutional layer has two output feature maps (real and imaginary channels of output). This filtering/denoising CNN is trained to reduce mean squared error between output and input channels via Adam optimizer.

We extract corresponding 64×64 patches from each real-world noisy training image and its thresholded coherence to train the coherence classification CNN. The method used to threshold the raw coherence is described in detail next.

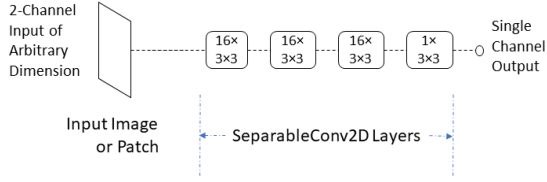

**Fig. 1**: Proposed CNN for InSAR coherence classification.

To train the CNN in Fig. 1, we first extract 7×7 patches and compute the raw coherence between the training noisy images and their filtered version output by the fully trained filtering network described earlier. Raw coherence is defined in Eq. 1:

$$\hat{\gamma} = \frac{\sum_{n,m} u_1(n,m) \cdot u_2^*(n,m) \cdot e^{-j\phi(n,m)}}{\sqrt{\sum_{n,m}|u_1(n,m)|^2 \sum_{n,m}|u_2(n,m)|^2}} \quad (1)$$

where pixel $n$ of interferogram $u_1$ and pixel $m$ of interferogram $u_2$ have the angle of separation $\Phi$, and the asterisk on top of $u_2$ denotes complex conjugate. A relatively large patch size (7×7) is used to reduce bias in raw coherence computation.

Next, for thresholding the raw coherence, we employ a Markov Random Field (MRF) based approach. Advantages of choosing MRF over simple histogram-based thresholding are: (a) spatial context is taken into consideration, and (b) smoothness of the thresholded regions (coherent and incoherent) can be adjusted by tuning a single parameter, as shown later. First, we initialize the pixel-wise estimates for the MRF using a fixed threshold. It was observed by running Otsu's global thresholding [35] on several raw coherence images that the average threshold is 0.6. Thus, we used 0.6 as the fixed threshold, such that pixel values above it are considered coherent (set to 1) and below it as incoherent (set to 0). Given these initial estimates $\{P_{ij} : 0$ or $1\}$, we try to find a solution $\{S_{ij} : 0$ or $1\}$ that minimizes Eq. 2

$$\Sigma_{ij}|P_{ij} - S_{ij}| + \Sigma_{ij}\Sigma_{kl}(\alpha * |S_{ij} - S_{(i+k)(j+l)}|) \quad (2)$$

where $\alpha$ is the smoothness parameter that needs to be manually tuned based on the results, and $|\:|$ is the $L_1$ norm. Also, $k$ and $l$ represent indices of pixel neighborhood. Since this is a binary classification (false:0 / true:1), exact solution for the expression in Eq. 2 was found via graph-cut optimization [36]. In Fig. 2, we show result of thresholding raw coherence by MRF for a sample training image. Black and white pixels represent 0 (low coherence) and 1 (high coherence) respectively. The value of $\alpha$ = 2.5 was experimentally found to produce this best result.

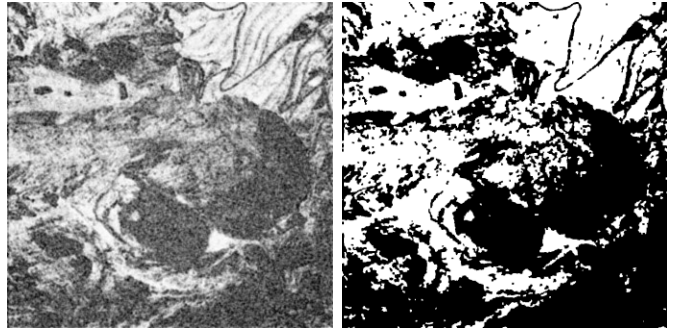

**Fig. 2**: Threshold raw 7×7 coherence via MRF and graph-cut.

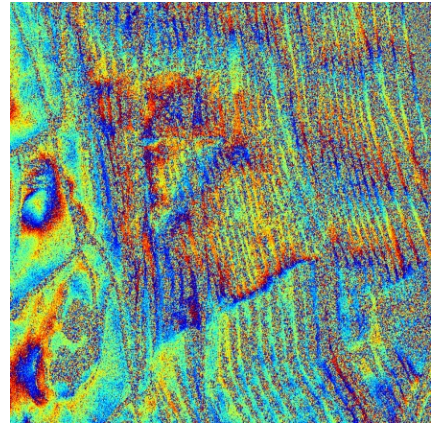

(a) Input Interferogram (Phase); Blue: -π; Red: +π

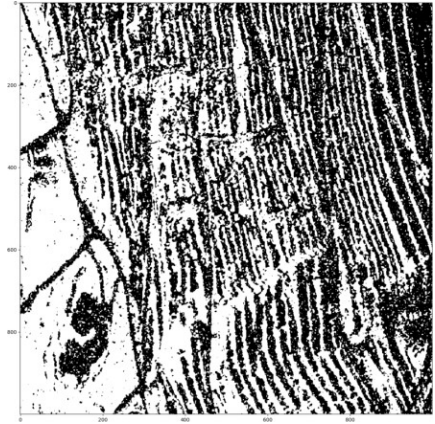

(b) Boxcar Coherence Output (Run time: **1.32 sec**)

Funding for this research was provided by MITACS and CARIC.

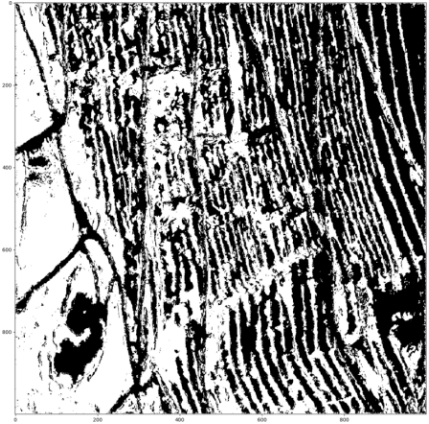
(c) NLInSAR Coherence Output [37] (Run time: **20.44 sec**)

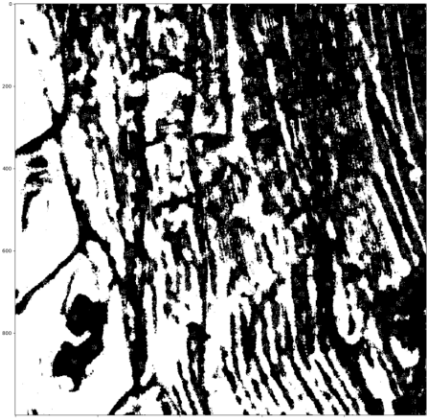
(d) NLSAR Coherence Output [38] (Run time: **11.49 sec**)

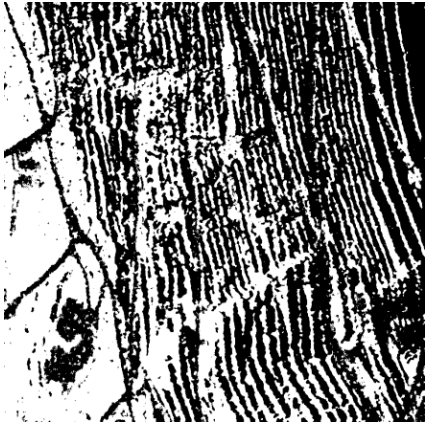
(e) Proposed Method Coherence Output (Run time: **0.67 sec**)

**Fig. 3**: Coherence classification of proposed vs. established methods: -ve (black: incoherent) and +ve (white: coherent).

## III. Results and Discussion

We implemented both networks (described earlier) using Keras with Tensorflow back-end. We trained the filtering CNN and coherence classifier CNN by extracting 500 60×60 sized and 64×64 sized patches respectively from each of 135 1000×1000 training interferograms. We used a large batch size of 100 to prevent overfitting. We started the training with learning rate $10^{-3}$ and halved it every 10 epochs for fast convergence. The CNNs converged after 50 and 100 epochs respectively. The trained CNNs were tested on 1000×1000 interferograms from a different geographic location, but using the interferogram itself as input, instead of just patches. Fig. 3 compares our method's coherence and run time performance with 3 existing methods.

Comparing outputs of all four methods against the input noisy interferogram in Fig 3a, we see that our method creates better, more accurate demarcation between coherent and incoherent regions, and lesser coherence misclassifications in completely incoherent areas than both Boxcar and NLSAR [37]. Our method does not have image border artefacts or false positives along borders of incoherent regions like NLInSAR [38]. A key advantage of our approach is: we can tune a single smoothness parameter as per the application requirements, while preparing training data to control the degree of smoothness of coherence classifications. Fig. 2 indicates that our method can be improved further by lowering bias in raw coherence. We multiply the input interferogram with its filtered version's complex conjugate to prevent the signals from biasing down coherence estimates, but we also found that wherever filtering fails to denoise properly, this process cancels out noise in noisy areas where some noise is still left in the filtered version. This drives up coherence. Thus, we can further reduce false positive classifications via better filtering. Also, the execution time of our classifier CNN is only 0.67 seconds, which is far less than other methods. All methods were implemented and executed in OpenCL 1.2 on a NVIDIA 1070 GPU with 8 GB GPU RAM.

To further validate our method, we simulated 100 clean ground truth interferograms with Gaussian bubbles, roads and buildings, added Gaussian noise, and input noisy versions to each method mentioned in Fig. 3, including ours. Performance score of each method's coherence classification with respect to ground truth classification for threshold = 0.6 are presented in Table I. Similarly, preprocessing accuracy, precision and recall were evaluated to be about 82, 80 and 95 percent respectively, and these confirmed the classifier CNN's training convergence.

TABLE I. Averaged Scores on 100 Simulated Interferograms

| Metric | *Boxcar* | *NLInSAR* | *NLSAR* | *Proposed* |
|---|---|---|---|---|
| **accuracy** | 0.8008 | 0.8273 | 0.4951 | 0.8425 |
| **precision** | 0.8248 | 0.8126 | 0.7389 | 0.8399 |
| **recall** | 0.8522 | 0.9265 | 0.2983 | 0.9107 |

Table I shows how our method outperforms others; NLInSAR recall is slightly higher, but we deduced from its output images that this is an anomaly, because it just overestimates coherence. Still, bias reduction in raw coherence computation and fine-tuning MRF thresholding to reduce resolution loss can generate even better results in our method, which is over 30 times faster than its closest rival, NLInSAR as per real and simulated data.

## IV. Conclusion

We propose a CNN-based coherence classifier for InSAR images. It creates far less misclassifications in incoherent areas than existing methods, and outperforms its closest competitor, NLInSAR by 30 times in run time. These show the capability of CNN-based learning for InSAR coherence classification.